\documentclass[showpacs,preprintnumbers,amsmath,amssymb,aps,twocolumn,superscriptaddress]{revtex4}

\usepackage{amsfonts}
\usepackage[dvips]{graphicx}
\usepackage{type1cm}
\usepackage{color}
\usepackage{bm}

\newcommand{\figref}[1]{Fig.~\ref{#1}}
\newcommand{\e}[1]{\text{e}^{#1}}

\newcommand{\tr}{\operatorname{Tr}}
\renewcommand{\vec}[1]{\mathbf{#1}}
\newcommand{\punc}[1]{\,#1}
\newcommand{\neweqnline}{\nonumber\\}
\newcommand{\eqnref}[1]{Eq.~\eqref{#1}}

\newcommand{\appref}[1]{App.~\ref{#1}}
\newcommand{\diffd}{\text d}

\begin{document}
\title{Itinerant ferromagnetism in a two-dimensional atomic gas}
\author{G.J.~Conduit}
\email{gjc29@cam.ac.uk}
\affiliation{Department of Condensed Matter Physics, Weizmann Institute of Science, Rehovot 76100, Israel}
\affiliation{Physics Department, Ben Gurion University, Beer Sheva 84105, Israel}

\date{\today}

\begin{abstract}
Motivated by the first experimental evidence of ferromagnetic behavior
in a three-dimensional ultracold atomic gas, we explore the possibility of itinerant
ferromagnetism in a trapped two-dimensional atomic gas. Firstly, we
develop a formalism that demonstrates how quantum fluctuations drive
the ferromagnetic reconstruction first order, and consider the
consequences of an imposed population imbalance. Secondly, we adapt
this formalism to elucidate the key experimental signatures of
ferromagnetism in a realistic trapped geometry.
\end{abstract}

\pacs{03.75.Ss, 75.20.En, 64.60.Kw, 75.45.+j}

\maketitle

\section{Introduction}

Itinerant ferromagnetism is a ubiquitous strongly correlated phase of
matter in the solid state. The theoretical study of itinerant
ferromagnetism dates back to the pioneering work of
Stoner~\cite{Stoner57} and Wohlfarth, which showed that ferromagnetism
emerges as repulsive pairwise interactions between electrons overcome
the kinetic energy penalty of polarization. Subsequent theoretical
work has determined that soft transverse magnetic fluctuations have
the potential to drive the ferromagnetic transition first order before
the quantum critical point is
reached~\cite{Abrikosov58,BelitzReview05,Duine05,Maslov06,Conduit08,Conduit09i}.
Phenomena consistent with a first order transition have been observed
in the solid state; though it is difficult to determine whether they
are due to soft magnetic fluctuations or the coupling of the magnetic
moment to phonon degrees of freedom. However,
Jo~\textit{et~al}.~\cite{Jo09} have recently presented the first
tentative evidence~\cite{LeBlanc09,Conduit09ii} of itinerant
ferromagnetism in an ultracold atomic gas. The cold atom gas is a
clean system in which to study ferromagnetism, completely devoid of
the interfering phonon degrees of freedom encountered in the solid
state, so gifts investigators with a valuable tool with which to
answer long-standing questions about solid state ferromagnets.
Furthermore, ultracold atoms experiments also present a unique opportunity
to explore fundamentally new physics associated with ferromagnetism
including the consequences of population imbalance~\cite{Conduit08}, a
conserved net magnetization~\cite{Berdnikov09}, the damping of
fluctuations by three-body loss~\cite{Conduit10}, spin
drag~\cite{Duine10}, and mass imbalance. Here we aim to take advantage
of the high levels of control investigators can exercise over the
external potential trapping the gas and turn to study ferromagnetism
in a two-dimensional thin film.

Itinerant ferromagnetism is difficult to observe in two dimensions in
the solid state~\cite{Chatterji01,Chang09}. However, it could be
realized in an ultracold atom gas by using counter-propagating lasers
to create one-dimensional potential which will lead to a stacked
two-dimensional gas. The system also offers investigators the
opportunity to study the possibility for a superconducting instability
to emerge near to the ferromagnetic phase
transition~\cite{Lohneysen07}. The two-dimensional system is of
particular interest in this case as it could shed light on high
temperature superconductivity where antiferromagnetism competes with
d-wave superconductivity to form the ground state. Here we adapt the
formalism introduced for the three-dimensional case~\cite{Conduit08}
to expose the contrasting behavior of the two-dimensional
ferromagnet. We develop a formalism that captures the effects of
transverse quantum fluctuations and explore how they renormalize the
effective interaction strength. We then address how population
imbalance modifies the behavior of the atomic gas before studying
ferromagnetic ordering in a trapped geometry.

\section{Field integral formalism}

It has been long established that quantum fluctuations in a
three-dimensional fermionic gas with repulsive interactions have the
potential to drive the ferromagnetic transition first
order~\cite{Abrikosov58,BelitzReview05,Duine05,Conduit08,Conduit09i}. To
investigate the impact of quantum fluctuations in a two-dimensional
fermionic gas we explore ferromagnetic reconstruction within the
setting of an atomic gas, adapting the phenomenology developed for the
three-dimensional case in Ref.~\cite{Conduit08}. We adopt this
formalism because unlike the Eliashberg theory~\cite{Maslov06} it
provides an exact expression for the free energy which then allows us
to make a prediction of the critical interaction strength for the
onset of ferromagnetism and study the atomic gas within a harmonic
well. Moreover, \emph{ab initio} Quantum Monte Carlo
calculations~\cite{Conduit09i,Pilati10} have recently been used to
verify the three-dimensional formalism, which should therefore provide
a solid foundation from which to study the two-dimensional
case. Although the atoms do not carry spin, we discriminate between
the two fermionic species with a pseudospin
$\sigma\in\{\uparrow,\downarrow\}$. The species cannot interconvert so
separate chemical potentials $\mu_{\sigma}$ tune the population
imbalance, which in turn pins the net polarization along the
pseudospin direction. However, when the spontaneous magnetization
formed exceeds the population imbalance, a nonzero in-plane
magnetization emerges. To study the potential for ferromagnetic
ordering we express the partition function as a fermionic coherent
state path integral $\mathcal{Z}=\tr\e{-\beta(\hat{H}-\mu
  \hat{N})}=\int\mathcal{D}\psi\,\e{-S}$ with the action
\begin{equation}
S=\int\sum_{\sigma=\{\uparrow,\downarrow\}} \bar\psi_\sigma \left(\partial_\tau
+\epsilon_{\hat{\vec{k}}}-\mu_{\sigma}\right)\psi_\sigma
+\int g\bar\psi_{\uparrow}\bar\psi_{\downarrow}\psi_{\downarrow}\psi_{\uparrow}\punc{.}
\label{stoneraction}
\end{equation}
Here $\int\equiv\int_0^\beta\diffd\tau\int\diffd^{2}r$ with reduced
temperature $\beta=1/k_{\text{B}}T$, and $\epsilon_{\hat{\bf k}}$
denotes the dispersion. As we wish to investigate two-dimensional
ferromagnetism we have constrained the spatial integral to a plane. A
two-dimensional atomic gas could be realized experimentally using
counter-propagating laser beams whose antinodes at half-wavelength
spacing $b$ will define stacked quasi-two-dimensional layers. Though
at finite temperature the ferromagnetic ordering is only marginally
stable, long-range order should be stabilized by the weak inter-plane
coupling~\cite{Berdnikov09}. The repulsive contact interaction
parameter $g=g\delta^{3}(\vec{r})$ that can be tuned with a Feshbach
resonance~\cite{Hertz} is linked to the s-wave scattering length $a$
in three dimensions through $g=\sqrt{2/\pi}a/b$~\cite{Bhaduri00}.
Unique to two dimensions, the interaction strength is independent of
density. This means that within a trapped geometry the entire atomic
gas will experience the same effective interaction strength and
therefore adopt the same polarization.

To develop an effective Landau theory of the magnetic transition,
Hertz introduced a scalar Hubbard-Stratonovich decoupling of the
two-body interaction term in the spin channel~\cite{Hertz}. However,
this form of decoupling neglects the potential impact of soft
transverse field fluctuations, which in three dimensions are
responsible for driving the second order transition first
order~\cite{Conduit08,Conduit09i}. Therefore, we will introduce a
general Hubbard-Stratonovich decoupling that incorporates fluctuations
in all of the spin $\bm\phi$ and charge $\rho$ sectors. Integrating
over the fermion degrees of freedom yields
$\mathcal{Z}=\int\e{-S}\mathcal{D}{\bm\phi}\mathcal{D}\rho$ with the
action
\begin{equation}
S=\int g(\phi^2-\rho^2)-\tr\ln[\partial_\tau+\epsilon_{\hat{\vec{k}}}-\mu_{\sigma_{\text{z}}}+g\rho
-g{\bm\sigma}\cdot{\bm\phi}]\punc{.}
\end{equation}
At this stage a saddle point analysis would determine the mean-field
values of $\rho$ and $\bm\phi$. However, quadratic fluctuations in
these auxiliary fields renormalize these equations. Therefore we
introduce the putative saddle point values $\rho_{0}$ for density and
$\vec{m}$ for magnetization, integrate out fluctuations in the
auxiliary fields, and finally minimize the energy to determine
$\rho_{0}$ and $\vec{m}$. It is also convenient to rotate the z-axis
from the quantization axis to lie along the direction of the saddle
point magnetization $\vec{m}$, with components labeled by
$s\in\{+,-\}$. After integrating over fluctuations in both the density
$\rho$ and magnetization channels $\bm\phi$ to Gaussian order, an
expansion of the action to second order in $g$ leads to
\begin{eqnarray}
{\cal Z}&=&\exp\Biggl[-\int g(m^2-\rho^2)+\tr\ln\hat{G}^{-1}\neweqnline
&-&\frac{g^{2}}{2}\tr(\hat{\Pi}_{+-}\hat{\Pi}_{-+}-\hat{\Pi}_{++}\hat{\Pi}_{--})\Biggr]\punc{,}
\end{eqnarray}
where we have defined the spin-dependent polarization operator
$\hat{\Pi}_{ss'}=\hat{G}_{s}\hat{G}_{s'}$, and
$\hat{G}^{-1}_{\pm}=\partial_\tau+\epsilon_{\hat{\vec{k}}}-\mu_{\pm}+g\rho_{0}\mp
g|\vec{m}|$. The contact interaction means that an unphysical
ultraviolet divergence arises from the term in the action that is
second order in $g$. To remove it we must affect the standard
regularization of the linear term $g(m^{2}-\rho^{2})$, setting
$g\mapsto\sqrt{2/\pi}a/b-2(\sqrt{2/\pi}a/b)^{2}A^{-1}
\sum_{\vec{k}_{3,4}}'(\epsilon_{{\bf k}_1}+\epsilon_{{\bf k}_2}-
\epsilon_{{\bf k}_3}-\epsilon_{{\bf k}_4} )^{-1}$~\cite{Pathria07},
where the prime indicates that the summation is subject to the
momentum conservation condition ${\bf k}_1+{\bf k}_2={\bf k}_3+{\bf
  k}_4$, and $A$ denotes the total area of one stacked layer.

Finally, after carrying out the remaining Matsubara summations, one obtains the following
expression for the free energy:
\begin{eqnarray}
\!\!\!\!\!\!\!\!\!\!\!\!&F&\!\!=\!\sum_{{\bf k},s=\pm}\epsilon_{\bf k}^{s}n_{s}\left(
\epsilon_{\bf k}\right)+\sqrt{\frac{2}{\pi}}\frac{a}{bA}N_{+}N_{-}\nonumber\\
\!\!\!\!\!\!\!\!\!\!\!\!&-&\!\!2\!\left(\!\!\sqrt{\!\frac{2}{\pi}}\!\frac{a}{bA}\!\!\right)^{2}\!\!\!\!{\sum_{{\bf k}_{1,2,3,4}}\!\!}'
\frac{n_{+}(\!\epsilon_{{\bf k}_1}\!)n_{-}(\!\epsilon_{{\bf k}_2}\!)[n_{+}(\!\epsilon_{{\bf k}_3}\!)\!+\!n_{-}(\!\epsilon_{{\bf k}_4}\!)]}
{\epsilon_{{\bf k}_1}+\epsilon_{{\bf k}_2}-
\epsilon_{{\bf k}_3}-\epsilon_{{\bf k}_4}}\punc{\!,}
\label{FreeEnergy}
\end{eqnarray}
where
$n_{s}(\epsilon)=1/[1+\e{\beta(\epsilon-\mu_{s}-s|\vec{m}|\sqrt{2/\pi}a/b)}]$
is the Fermi distribution, and
$N_{s}=\sum_{\vec{k}}n_{s}(\epsilon_{\vec{k}})$. To evaluate the final
nine-dimensional integral in \eqnref{FreeEnergy} numerically we employ
the re-parameterization outlined in
\appref{sec:ComputationalAnalysisOfMomentumSpaceIntegral} to reduce it
to a four-dimensional integral. Moreover, as we are interested in
searching for extrema in the free energy with changing polarization we
can differentiate our expression with respect to magnetization, which
at zero temperature further reduces the integral to just three
dimensions.

To highlight the potential importance of fluctuation corrections we
briefly study the contribution to the energy from particle-hole
excitations around momentum $2k_{\text{F}}$. At zero temperature a
non-analytic contribution to the free energy of the form
$|m|^{3}\ln{m^{2}}$ emerges. The same non-analyticity was found
diagrammatically in Refs.~\cite{Maslov06}. The formation of a finite
magnetization increases the phase-space available for the formation of
virtual intermediate pairs of particle-hole pairs, and this phase
space enhancement donates a non-analytic term to the free energy
giving the transition the potential for first order character. In the
next section we study the effect that this non-analyticity has
on the phase diagram.

\section{Phase behavior}

With the formal development of the theory complete we will now apply
the formalism to explore the implications of ferromagnetism in the
two-dimensional atomic Fermi gas, and critically compare the results
with the three-dimensional case~\cite{Conduit08}.  Before we study the
phase diagram of the fluctuation corrected free energy, to make
contact with the conventional Stoner theory we first consider the
result of a direct saddle point approximation scheme in which the
second order term in the free energy is neglected. In this
approximation at zero temperature the free energy is
$F=(1+a/b\sqrt{2}\pi^{3/2})\mu^{2}/2\pi+(1-a/b\sqrt{2}\pi^{3/2})m^{2}/2\pi\mu^{2}$.
This expression is exact, and with magnetization featuring only as the
lowest available term in a Landau expansion its analysis is
straightforward. For $a<\sqrt{2}\pi^{3/2}b\approx7.874b$ this model
predicts that the gas is paramagnetic, whereas for
$a>\sqrt{2}\pi^{3/2}b$ the system is fully polarized, a scenario that
remains unaltered with the introduction of population imbalance. An
immediate corollary is that the spontaneous magnetization formed is
independent of the local density, which also holds true when
fluctuation corrections are taken into account. Therefore, within a
trap, the entire atomic gas adopts the same polarization.

\begin{figure}
 \centerline{\resizebox{0.85\linewidth}{!}{\includegraphics{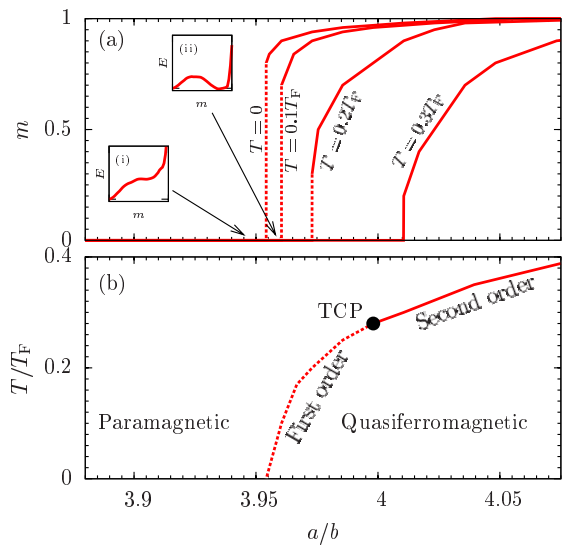}}}
 \caption{(Color online) (a) The growth of magnetization $m$ with
   scattering length for different temperatures. The inset figures
   show the energy landscape with magnetization for $T=0$ either side
   of the first order transition. (b) the phase diagram of temperature
   with scattering length shows the first order (dashed red line) and
   second order (solid red line) (quasi)ferromagnetic ordering from the
   paramagnetic phase.}
 \label{fig:CanonicalEnsemble_Order}
\end{figure}

\begin{figure}
 \centerline{\resizebox{0.8\linewidth}{!}{\includegraphics{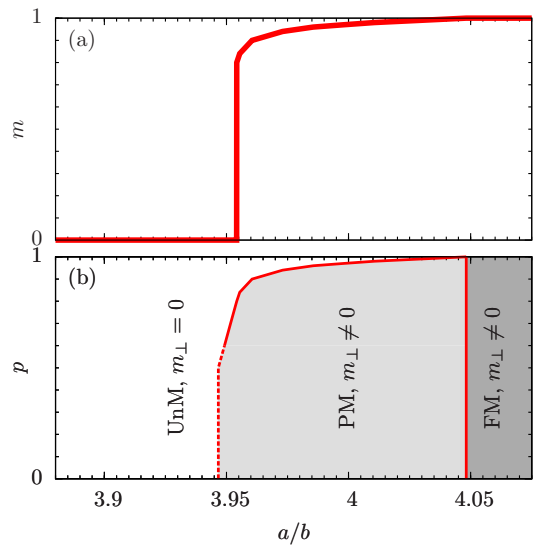}}}
 \caption{(Color online) (a) The growth of magnetization $m$ with
   scattering length at $T=0$. (b) the $T=0$ phase diagram for imposed
   population imbalance $p$ with scattering length $a/b$ shows the
   first order (dashed red line) and second order (solid red line)
   ferromagnetic ordering from the unmagnetized (UnM) to the partially
   magnetized (PM) and fully magnetized (FM) regions, here
   unmagnetized refers to having no in-plane magnetization.}
 \label{fig:CanonicalEnsemble_Pop_Imb}
\end{figure}

Having studied the mean-field limit, we now consider the repercussions
of fluctuation corrections on the behavior of the magnetization.  To
orient our discussion, we first consider a gas with equal populations
of up and down-spin atoms. As shown in
\figref{fig:CanonicalEnsemble_Order}(ai) at scattering lengths below
$a\approx3.945b$ the energy profile possesses a single minimum at zero
magnetization. With rising interaction strength a second minimum in
the energy landscape develops at $m\approx0.6$, which, with rising
scattering length, deepens in
\figref{fig:CanonicalEnsemble_Order}(aii) to become the global minimum
at $a\approx3.953b$ and $m\approx0.8$. At this scattering length the
system undergoes a first order transition from $m=0$ into the
polarized regime with $m\approx0.8$. As shown in
\figref{fig:CanonicalEnsemble_Order}(a) with a further increase in the
interaction strength the magnetization saturates at a scattering
length $a\approx4.048b$.  Fluctuation corrections have had significant
impact: they have driven the ferromagnetic transition to a
significantly weaker interaction strength ($a\approx3.953b$) compared
to the mean-field case ($a\approx7.874b$). At this weaker interaction
strength the $m^{2}$ term in the free energy has a positive
coefficient, and the ordering is driven by the non-analytic
$|m|^{3}\ln m^{2}$ term. The abetment of the transition by fluctuation
corrections and reduction in interaction strength at which
ferromagnetism is seen is common to both the two and three-dimensional
cases, though in two dimensions the transition is immediately to full
polarization at mean-field level and fluctuation corrections drive a
first order transition at a weaker interaction strength.

We now turn
to address the behavior of the phase transition at finite temperature
in \figref{fig:CanonicalEnsemble_Order}(a). Increasing temperature
dulls the fluctuation corrections and the scattering length of the
first order transition rises and the magnetization following the
transition is reduced. \figref{fig:CanonicalEnsemble_Order}(b) shows
that at $T\approx0.28T_{\text{F}}$ a tricritical point emerges and the
system reverts to second order behavior. The Mermin-Wagner-Hohenberg
theorem~\cite{Mermin66} states that although an ordered phase can
exist in two dimensions at zero temperature, at any finite temperature
fluctuations will destroy long range correlations in the system, and the
state will be characterized by exponentially decaying correlation functions. Therefore we denote
the ferromagnetic state as a ``quasiferromagnet''
(ferromagnet with fluctuating polarization direction). However, since the two-dimensional gas is
experimentally realized in a series of disks, each one can couple to its neighbors and 
tunneling should stabilize the phase~\cite{Petrov00,Berdnikov09}. Furthermore,
the Mermin-Wagner-Hohenberg theorem is valid only in
the thermodynamic limit and does not apply to finite-sized
systems. For a two-dimensional Bose gas with attractive interactions
it has been shown that a potential trap restricts the system and
stabilizes a quasi-Bose Einstein condensate~\cite{Andersen02}. In a similar way the harmonic trap
should stabilize a ferromagnetic phase. So far we have focused on how equilibrium properties
can stabilize the ferromagnetic phase, however, there are also non-equilibrium aspects to consider.
Within the current experimental realization of cold atom gas ferromagnetism
three-body losses necessitate that the experiment be performed out of equilibrium.
Following a quench small ferromagnetic domains
are formed~\cite{Babadi09} which then grow steadily~\cite{Conduit09ii}.
The final size of these ferromagnetic domains
$\sim6/k_{\text{F}}$~\cite{Conduit09ii} at $T=0.1T_{\text{F}}$ and $k_{\text{F}}a=2$ is
small compared to the length-scale of the thermal fluctuations given by
$a\exp[2\pi(2k_{\text{F}}a/\pi-1)T_{\text{F}}/T]\approx10^{7}/k_{\text{F}}$~\cite{Auerbach94}. This
means that at sufficiently low temperature fluctuations will not disrupt the ferromagnetic state and so
in experiments a true ferromagnetic phase should be observed
as shown in \figref{fig:CanonicalEnsemble_Order}.

Having addressed the situation without population imbalance, we now
consider how a fixed spin population imbalance influences the phase
diagram. The two constituent species cannot interconvert so an initial
population imbalance is maintained by the difference in their chemical
potentials. However, if energetically favorable, the gas can become
more polarized either by phase separation or the development of an
in-plane magnetic moment. As shown in
\figref{fig:CanonicalEnsemble_Order}(ai and aii), at weak interactions
such that $a\lessapprox3.945b$, the energy monotonically increases with
magnetization, but the magnetization remains pinned to the minimum
value defined by the population imbalance $p$. However, with rising
interaction strength a second minimum develops in the free energy
landscape from $a\approx3.945b$ and $m\approx0.6$. If that
magnetization exceeds the population imbalance, then as shown in
\figref{fig:CanonicalEnsemble_Pop_Imb}(b) the system will phase
separate between this minimum and that at zero magnetization, with
relative fractions governed by the Maxwell construction. When the
emerging minimum becomes the global minimum at $m\approx0.8$, then
gases with a lower population imbalance enter this global minimum with
an appropriate in-plane magnetic moment. As the magnetization of the
minimum rises it envelops systems with higher population imbalance,
and tracks the magnetization curve shown in
\figref{fig:CanonicalEnsemble_Pop_Imb}(a) until it reaches full
polarization at $a/b\approx4.048$. Like the three-dimensional case,
the population imbalance renders the characteristic interaction
strength of the transition to be almost constant up to an imposed
population imbalance of $p\approx0.8$, which could be a key
experimental signature of first order behavior.

\section{Trapped geometry}

Having addressed the phase behavior of a uniform system, to make
contact with the experiment we now turn to address the atomic gas
trapped within the spherical potential $V(\vec{r})=\omega
r^{2}/2$. Following the program developed in
Refs.~\cite{LeBlanc09,Conduit09ii,Berdnikov09} we aim to minimize the
free energy within the local density approximation using the kernel
$f(\vec{r})=F(\vec{r})+V(\vec{r})[n_{+}(\vec{r})+n_{-}(\vec{r})]-\gamma_{+}n_{+}(\vec{r})+\gamma_{-}n_{-}(\vec{r})$,
here $F(\vec{r})$ denotes the energy kernel \eqnref{FreeEnergy}
evaluated with the local chemical potential at $\vec{r}$. The
Lagrange multipliers $\gamma_{\pm}$ enforce the constraints of
constant number of atoms imposed by the trap geometry
$N_{\text{tot}}=\int [n_{+}(\vec{r})+n_{-}(\vec{r})]\diffd^{2}r$ and
population imbalance $p\leq\int
[n_{+}(\vec{r})-n_{-}(\vec{r})]\diffd^{2}r/N_{\text{tot}}$; without
loss of generality we assume that $p\ge0$ and therefore
$\gamma_{+}\ge\gamma_{-}$. To study the effects of spatial density
variations we invoke a local density approximation that enables the
variational minimization $\delta f/\delta n_{s}(\vec{r})$ and yields
the simultaneous equations for the effective local chemical potentials
$\mu_{\pm}(\vec{r})$ for the species in the rotated spin basis
%
\begin{eqnarray}
 &&\!\!\!\!\!\!\!\!\mu_{\pm}(\vec{r})\!=\!\gamma_{\pm}\!-\!V\!(\vec{r})\!-\!\sqrt{\!\frac{2}{\pi}}\!\frac{a}{bA}n_{\mp}(\vec{r})\!+\!2\!\!\left[\!\sqrt{\!\frac{2}{\pi}}\!\frac{a}{bA}\!\right]^{2}\!\!\!\!{\sum_{\vec{k}_{1,2,3,4}}\!\!}'\!\!\!n_{\mp}(\!\epsilon_{\vec{k}_{2}}\!)\neweqnline
&&\!\!\!\!\!\!\!\!\times\!\frac{n_{\pm}(\epsilon_{\vec{k}_{1}})\delta(\epsilon_{\vec{k}_{3}}\!\!-\!\mu_{\pm})\!+\![n_{\pm}(\epsilon_{\vec{k}_{3}})\!+\!n_{\mp}(\epsilon_{\vec{k}_{4}})]\delta(\epsilon_{\vec{k}_{1}}\!\!-\!\mu_{\pm})}{\epsilon_{\vec{k}_{1}}+\epsilon_{\vec{k}_{2}}-\epsilon_{\vec{k}_{3}}-\epsilon_{\vec{k}_{4}}}\!\punc{.}
 \label{eqn:EulerLagrangeEqn}
\end{eqnarray}
These equations can be understood as having been constructed out of
three orders of perturbation theory. The lowest, independent of the
scattering length $a$, corresponds to the Thomas-Fermi approximation
within the confining potential, the term first order in $a$ introduces
the mean-field energy penalty of the interaction whereas the second
order term introduces the energy associated with magnetic quantum
fluctuations. The detailed study of the uniform system revealed that
the polarization depends only on the interaction strength and not
spatial density variations, meaning that the ratio
$n_{+}(\vec{r})/n_{-}(\vec{r})$ and therefore
$\mu_{+}(\vec{r})/\mu_{-}(\vec{r})$ is constant across the trap.
Therefore, the two equations reduce to just one that is solved by
iteration.

\subsection{Heuristic observations}

We first address what can be determined about the behavior of the
atomic gas heuristically before presenting the results of the full
solution of \eqnref{eqn:EulerLagrangeEqn}. To develop our intuition we
focus on perhaps the most physical quantity that can be measured by
experiment, namely the cloud size. To start the analysis we consider
the non-interacting limit $a=0$ where the system is unpolarized and
the effective chemical potentials given by
\eqnref{eqn:EulerLagrangeEqn} follow the familiar Thomas-Fermi
form. The root mean square (RMS) radius would increase with population
imbalance as $[1+(\frac{1-p}{1+p})^{3/2}]^{1/2}/\sqrt{2}$ due to the
increasing Fermi degeneracy pressure. With weak interactions
$a\ll\sqrt{2}\pi^{3/2}b$ we need consider
\eqnref{eqn:EulerLagrangeEqn} only to first order in $a$ which yields
$\mu_{\pm}(\vec{r})=\gamma_{\pm}-V(\vec{r})-a\max[\gamma_{\mp}-V(\vec{r}),0]/2^{1/2}\pi^{3/2}b$. The
first order term reduces the effective chemical potential so to
conserve the total number of trapped atoms we renormalize the Lagrange
multipliers upwards from the Thomas Fermi value by a factor of
$1+a(1-p)/(2\pi)^{3/2}b(1+p)$ for the majority spin species and
$1+a[2-(\frac{1-p}{1+p})^{1/2}]/(2\pi)^{3/2}b$ for the minority spin
species. This reduction in the effective chemical potential and
corresponding fall in local density can be understood in terms of an
increase in the local pressure within the cloud due to the repulsive
interactions between the atoms. This pressure inflates the cloud
causing the RMS radius to rise through a factor of
$1+a(1-p)(3\sqrt{1+p}-\sqrt{1-p})/2^{5/2}\pi^{3/2}b[(1+p)^{3/2}+(1-p)^{3/2}]$. Having
analyzed the weakly interacting regime it is natural to also examine
the strongly interacting limit. Here the atomic gas is fully polarized
so $\mu_{-}=0$ and $\mu_{+}(\vec{r})=\gamma_{+}-V(\vec{r})$, meaning
that the system is firmly in the Thomas Fermi regime. We again require
that the number of particles is conserved which sets the majority spin
Lagrange multiplier to rescale by a factor of $2^{1/2}$ from its
original value if there were no population imbalance. Consequentially
the enhanced Fermi degeneracy pressure dilates the RMS radius of a
cloud with zero population imbalance by a factor of $2^{1/4}$. The key
limits of weak and strong interactions hold true whatever the true
theory of ferromagnetism so provide two valuable handles for potential
experiments.

\subsection{Exact analysis of trapped behavior}

\begin{figure}
 \centerline{\resizebox{0.95\linewidth}{!}{\includegraphics{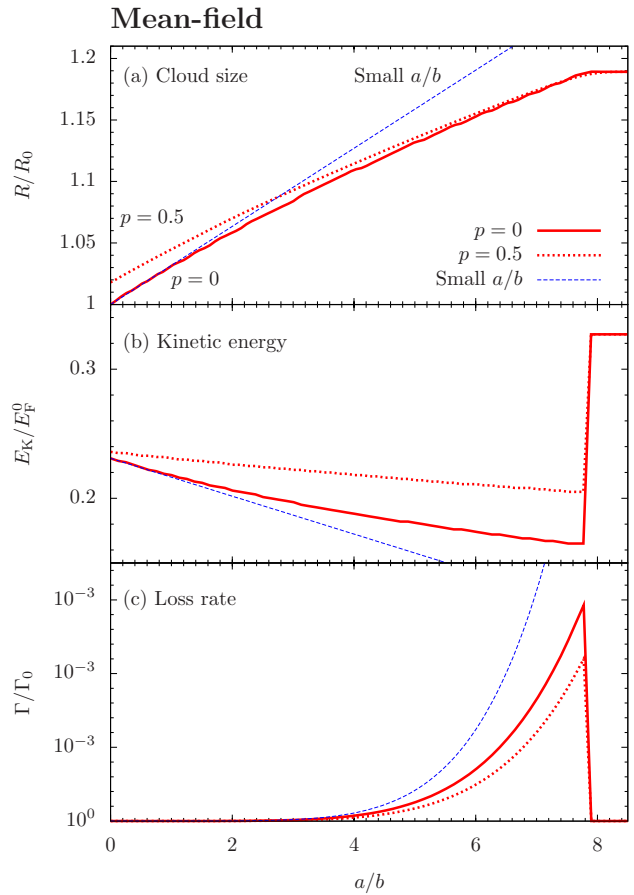}}}
 \caption{(Color online) The variation of (a) cloud size, (b) kinetic
   energy, and (c) atom loss rate on ferromagnetic ordering with
   increasing scattering length $a/b$ for the orthodox Stoner
   mean-field theory case. The thin blue dashed line highlights the small $a/b$
   behavior. The solid lines are at zero population imbalance whereas
   the dotted line is with an imposed population imbalance of $0.5$.}
 \label{fig:TrapBehavior_MF}
\end{figure}

\begin{figure}
 \centerline{\resizebox{0.95\linewidth}{!}{\includegraphics{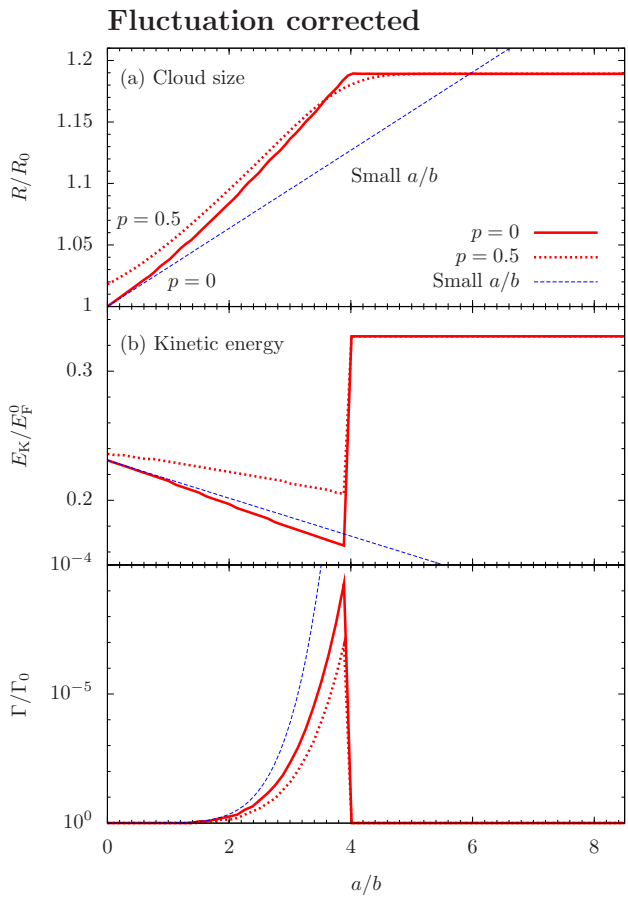}}}
 \caption{(Color online) The variation of (a) cloud size, (b) kinetic
   energy, and (c) atom loss rate on ferromagnetic ordering with
   increasing scattering length $a/b$ when fluctuation corrections are
   taken into account. The thin blue dashed line highlights the small $a/b$
   behavior. The solid lines are at zero population imbalance whereas
   the dotted line is with an imposed population imbalance of $0.5$.}
 \label{fig:TrapBehavior_FC}
\end{figure}

Having completed the overview of the trapped behavior we now turn to
consider the ramifications of fluctuation corrections and
self-consistently solve \eqnref{eqn:EulerLagrangeEqn} for the chemical
potentials $\mu_{\pm}$. We then integrate over the trap to extract the
full behavior of the experimental observables, namely cloud size,
kinetic energy, and three-body loss rate, which for the mean-field
limit are shown in \figref{fig:TrapBehavior_MF}. The same calculation
repeated for fluctuation corrections is shown in
\figref{fig:TrapBehavior_FC}. A useful reference throughout will be
the complementary analysis in three
dimensions~\cite{LeBlanc09,Conduit09ii}. The orthodox Stoner
mean-field theory predicts that at the onset of ferromagnetic ordering
the system immediately fully polarizes across the entire trap at
$a/b=\sqrt{2}\pi^{3/2}\approx7.874$, whereas fluctuation corrections
allow the cloud to adopt partial polarization over the window of
scattering lengths $3.953\lessapprox a/b\lessapprox4.048$. In
two-dimensions as the entire gas polarizes at the same interaction
strength striking features emerge at these respective interaction
strengths. Current experiments~\cite{Jo09} can probe scattering
lengths to $\sim10\%$ accuracy, therefore in current experiments the
fluctuation corrected transition will also appear to immediately give
complete polarization.

To develop our intuition we first examine the projected \emph{cloud
  size}. In the mean-field limit (\figref{fig:TrapBehavior_MF}(a))
with weak interactions the RMS radius grows linearly with scattering
length as the atoms repel each other within the trap. The radius grows
following the universal scaling described above. Population imbalance
causes the cloud to have an initially larger radius due to the
increased Fermi degeneracy pressure. The entire cloud becomes fully
polarized at the same scattering length, $a/b=\sqrt{2}\pi^{3/2}$, and
at this point the cloud size immediately adopts its final inflated
radius $R/R_{0}=2^{1/4}$, maintained by Fermi degeneracy
pressure. This is in contrast to the three-dimensional
case~\cite{LeBlanc09,Conduit09ii} in which the transition takes place
over a range of interaction strengths, thus making the transition less
distinct. \figref{fig:TrapBehavior_FC}(a) shows that fluctuation
corrections drive the cloud expansion faster, causing it to dilate
rapidly. In contrast to the three-dimensional
case~\cite{LeBlanc09,Conduit09ii}, this pressure cannot drive the
cloud to grow larger than the fully polarized size
$2^{1/4}R_{0}^{\text{RMS}}$. As the interaction strength is unaffected
by the density of atoms, the transition occurs at the same scattering length
$a\approx3.954b$ seen in the uniform case.

The \emph{total kinetic energy} is probed experimentally by releasing
the atoms from the trap and imaging them following a ballistic
expansion. Starting from the mean-field analysis in
\figref{fig:TrapBehavior_MF}(b), at zero interactions an initial
population imbalance increases the kinetic energy due to the enlarged
majority spin Fermi surface by a factor of
$[1+(\frac{1-p}{1+p})^{3/2}]/2$. The weak interactions dilate the
cloud, causing local density and kinetic energy to fall with the
universal scaling
$1-a(1-p)(3\sqrt{1+p}+\sqrt{1-p})/2^{5/2}\pi^{3/2}b[(1+p)^{3/2}+(1-p)^{3/2}]$.
When the scattering length is increased beyond $a/b=\sqrt{2}\pi^{3/2}$
the entire gas becomes ferromagnetic and the atoms all enter the same
Fermi surface. This Fermi surface is inflated and the kinetic energy
plateaus at the final value that is $2^{1/2}$ times that for the
non-interacting gas. When fluctuation corrections are taken into
account one recovers the variation of kinetic energy shown in
\figref{fig:TrapBehavior_FC}(b). The fluctuations drive the transition
to take place at a reduced interaction strength $a\approx3.954b$ seen
in the uniform case.

The \emph{atom loss rate} due to three-body recombination is
$\Gamma=\Gamma_{0}(a/b)^{6}\int
n_{+}(\vec{r})n_{-}(\vec{r})[n_{+}(\vec{r})+n_{-}(\vec{r})]\diffd^{2}r$~\cite{Petrov03}. In
the recent experiment~\cite{Jo09} the three-body loss was significant
and forced the experiment to be performed rapidly and
out of equilibrium, and here we study the situation in
two-dimensions. We start by examining the mean-field limit in
\figref{fig:TrapBehavior_MF}(c), which shows the three-body loss
integrated over the entire trap. At weak interaction strengths the
loss rate rises rapidly as
$\Gamma=\Gamma_{0}(a/b)^{6}\mu^{4}(1-p^{2})/8\pi^{2}\omega$. At a
scattering length $a/b=\sqrt{2}\pi^{3/2}$ the gas across the entire
trap becomes fully polarized so $n_{-}=0$ and therefore the three-body
loss is completely cut off. This immediate elimination of loss
contrasts the three-dimensional case where loss remains until high
interaction strengths, where it forces the experiment out of
equilibrium~\cite{Conduit09ii}, and also renormalizes the effective
interaction strength~\cite{Conduit10}. \figref{fig:TrapBehavior_MF}
highlights how these effects are reduced in the two-dimensional case
which could aid with the positive identification of the ferromagnetic
phase. It can also be seen that population imbalance reduces atom loss
primarily through reduction of the $n_{+}(\vec{r})n_{-}(\vec{r})$
term. Having studied the mean-field limit we now look at the
impact of fluctuation corrections on three-body loss in
\figref{fig:TrapBehavior_FC}(c). The fluctuation corrections drive the
ferromagnetic transition to take place at a reduced scattering length
of $a/b=3.954$. This in turn means that the peak three-body loss
$(\propto a_{\text{crit}}^{6})$ is significantly reduced. This fall in
loss rate will mean that an experiment searching for signatures of
ferromagnetism can be performed nearer to the equilibrium regime which
should yield clearer results.

\section{Discussion}

In conclusion, on the repulsive side of the Feshbach resonance
coupling of transverse magnetic fluctuations drives ferromagnetic
ordering first order. We studied the specific variation of three
experimental signatures of ferromagnetism: cloud size, release energy,
and atom loss rate. The formalism highlighted the benefits of studying
ferromagnetism in two rather than three dimensions. In two-dimensions
the effective interaction strength is independent of density and
therefore radius in the harmonic well. As the interaction strength is
ramped upwards the entire gas will enter into the ferromagnetic phase
at the same Feshbach field, whereas in three-dimensions the gas first
enters the ferromagnetic state at the center. Therefore the signatures
of the ferromagnetic phase are enhanced in two-dimensions, which
should aid the exact characterization of the state. At weak
interactions these observables displayed universal scaling, and the
variation with an imposed population imbalance was also considered.

One intriguing possibility opened up by the new formalism developed to
study fluctuation corrections is ferromagnetic reconstruction into a
spin textured state, in a matter analogous to the FFLO state in
superconductors. This has already been shown to be possible in three
dimensions~\cite{Conduit09i} and, with enhanced Fermi surface nesting
in two dimensions, poses an interesting direction for future research.

The author thanks Ehud Altman, Andrew Green, Gyu-Boong Jo, Wolfgang
Ketterle, Ben Simons, and Joseph Thywissen for useful discussions.
The author acknowledges the financial support of the Royal Commission
for the Exhibition of 1851 and the Kreitman Foundation.

\clearpage

\appendix
\section{Computing the momentum space integral}\label{sec:ComputationalAnalysisOfMomentumSpaceIntegral}

\begin{figure}
 \centerline{\resizebox{0.9\linewidth}{!}{\includegraphics{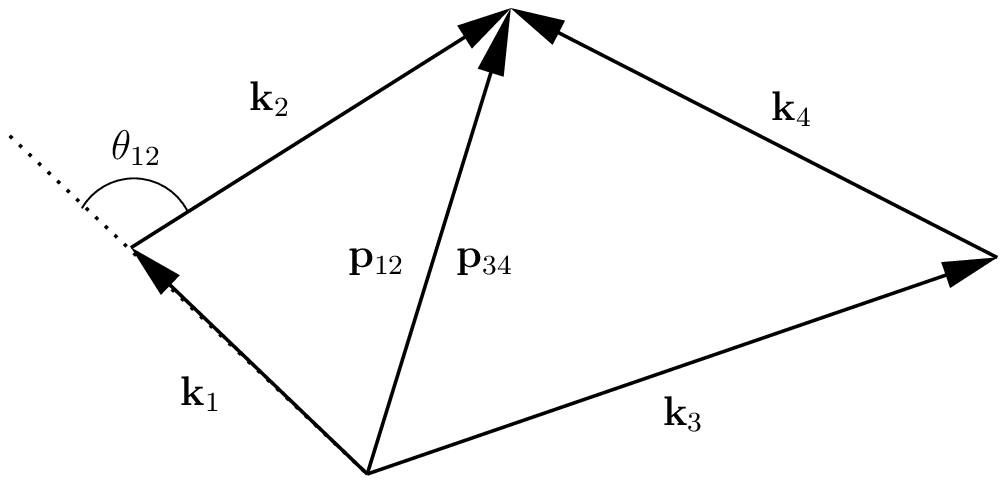}}}
 \caption{The re-parameterization of the momenta
   $\vec{k}_{1,2,3,4}$. $\theta_{12}$ represents the angle between
   $\vec{k}_{1}$ and $\vec{k}_{2}$. The two momenta
   $\vec{p}_{12}=\vec{k}_{1}+\vec{k}_{2}$ and
   $\vec{p}_{34}=\vec{k}_{3}+\vec{k}_{4}$ are constrained to be equal,
   $\vec{p}_{12}=\vec{p}_{34}$, by the Dirac delta function in
   \eqnref{eqn:GeneralQuadkIntegralToBeReParameterised}.}
 \label{fig:FourthOrderMomSpaceInt}
\end{figure}

An important integral \eqnref{FreeEnergy} encountered in this paper has the
form
\begin{equation}
 \label{eqn:GeneralQuadkIntegralToBeReParameterised}
 \iiiint\!\!\diffd\vec{k}_{1}\diffd\vec{k}_{2}\diffd\vec{k}_{3}\diffd\vec{k}_{4}F(k_{1},k_{2},k_{3},k_{4})\delta(\vec{k}_{1}+\vec{k}_{2}-\vec{k}_{3}-\vec{k}_{4})\punc{.}
\end{equation}
To evaluate this integral one could substitute
$\vec{k}_{4}=\vec{k}_{1}+\vec{k}_{2}-\vec{k}_{3}$, and then integrate over
the three parameters representing the lengths of vectors $\vec{k}_{1}$,
$\vec{k}_{2}$, and $\vec{k}_{3}$, and a minimum of three relative angles
between these vectors, giving a total of six integration
parameters. However, since numerical integration generally becomes
prohibitive with increasing number of dimensions we outline a scheme that
takes advantage of the fact that the function $F$ depends only on the
magnitude of the momenta to perform the angular integrals and leave a
numerical integral over just the four vector lengths. A similar
scheme has been developed in the three-dimensional case~\cite{Conduit08}.

The integral is re-parameterized according to
\figref{fig:FourthOrderMomSpaceInt}. The angular integral associated with
vectors $\vec{k}_{1}$ and $\vec{k}_{2}$ is $\int_{0}^{2\pi}2\pi
k_{1}k_{2}\diffd\theta_{12}$, where $\theta_{12}$ is the angle between $\vec{k}_{1}$
and $\vec{k}_{2}$. We now change the variable of the angular
integral over $\theta_{12}$ to the vector
$\vec{p}_{12}=\vec{k}_{1}+\vec{k}_{2}$ through the relationship
$\cos\theta_{12}=(k_{1}^{2}+k_{2}^{2}-p_{12}^{2})/2k_{1}k_{2}$ and so
$\int_{0}^{2\pi}\diffd\theta_{12}=\int_{|k_{1}-k_{2}|}^{k_{1}+k_{2}}\diffd
p_{12}8\pi
k_{1}k_{2}p_{12}[4k_{1}^{2}k_{2}^{2}-(k_{1}^{2}+k_{2}^{2}-p_{12}^{2})^{2}]^{-1/2}$. This
expression, and an analogous one in
$\vec{p}_{34}=\vec{k}_{3}+\vec{k}_{4}$, allows us to rewrite the original
integral \eqnref{eqn:GeneralQuadkIntegralToBeReParameterised} in terms of
the parameters $p_{12}$ and $p_{34}$. The momentum conservation requirement
is imposed by $\delta(\vec{k}_{1}+\vec{k}_{2}-\vec{k}_{3}-\vec{k}_{4})$
which now introduces a new conservation law
$\delta(\vec{p}_{12}-\vec{p}_{34})$. This sets the two integration
parameters equal, $p_{12}=p_{34}=p$, so there is just one integral over
parameter $p$ remaining, and since the delta function constrains 
the angle between $\vec{p}_{12}$ and $\vec{p}_{34}$ we must also
divide by the phase space associated with the angular integration of 
$2\pi p$. We then obtain
\begin{eqnarray*}
 &&\!\!\!\!\!\!32\pi\iiiint\diffd k_{1}\diffd k_{2}\diffd k_{3}\diffd k_{4}\int_{\max(|k_{1}-k_{2}|,|k_{3}-k_{4}|)}^{\min(k_{1}+k_{2},k_{3}+k_{4})}\diffd p\times\nonumber\\
 &&\!\!\!\!\!\!\frac{F(k_{1},k_{2},k_{3},k_{4})k_{1}k_{2}k_{3}k_{4}p}{\sqrt{4k_{1}^{2}k_{2}^{2}-(k_{1}^{2}+k_{2}^{2}-p^{2})^{2}}\sqrt{4k_{3}^{2}k_{4}^{2}-(k_{3}^{2}+k_{4}^{2}-p^{2})^{2}}}\punc{.}
\end{eqnarray*}
Finally, we note that the integral over variable $p$ is Carlson's standard
elliptic integral of the first kind, which we denote by $R_{\text{F}}$. This
yields the final result
\begin{eqnarray*}
 &&\!\!\!\!\!\!32\pi\iiiint\diffd k_{1}\diffd k_{2}\diffd k_{3}\diffd
  k_{4}F(k_{1},k_{2},k_{3},k_{4})\times\nonumber\\
 &&\!\!\!\!\!\!\Theta(k_{1}+k_{2}-|k_{3}-k_{4}|)\Theta(k_{3}+k_{4}-|k_{1}-k_{2}|)\times\nonumber\\
 &&\!\!\!\!\!\!\frac{R_{\text{F}}\left(0,1+\left|\frac{[(k_{1}+k_{2})^2-(k_{3}-k_{4})^2][(k_{1}-k_{2})^2-(k_{3}+k_{4})^2]}{[(k_{1}+k_{2})^2-(k_{3}+k_{4})^2][(k_{1}-k_{2})^2-(k_{3}-k_{4})^2]}\right|,1\right)}{\sqrt{|[(k_{1}-k_{2})^2-(k_{3}-k_{4})^2][(k_{1}+k_{2})^2-(k_{3}+k_{4})^2]|}}
 \punc{.}
\end{eqnarray*}
The term introduced to compensate for the angular integrals can be
efficiently computed by a suitable numerical library. This four-dimensional
integral is now better suited to computational evaluation than the
six-dimensional form of the original expression
\eqnref{eqn:GeneralQuadkIntegralToBeReParameterised}.

\end{document}